\documentclass[11pt]{article}

\usepackage[T1]{fontenc}
\usepackage[utf8]{inputenc}
\usepackage{lmodern}
\usepackage[a4paper,margin=1in]{geometry}
\usepackage{microtype}
\usepackage{graphicx}
\usepackage{amsmath,amssymb}
\usepackage{centernot}
\usepackage{booktabs}
\usepackage{multirow}
\usepackage{tabularx}
\usepackage{longtable}
\usepackage{array}
\usepackage[table]{xcolor}
\usepackage{enumitem}
\usepackage{caption}
\usepackage{subcaption}
\usepackage{float}
\usepackage{placeins}
\usepackage{listings}
\usepackage{url}
\usepackage[colorlinks=true,linkcolor=blue,citecolor=blue,urlcolor=blue]{hyperref}
\usepackage[nameinlink,noabbrev]{cleveref}
\hypersetup{
  pdftitle={X-LocVul: An Evidence-Propagation Audit of Cascaded C/C++ Vulnerability Detection, Localization, and Explanation},
  pdfauthor={Hoang-Phuong-Linh Le; Quyet-Thang Huynh},
  pdfsubject={Preprint on evidence propagation in vulnerability-analysis cascades},
  pdfkeywords={vulnerability detection, vulnerability localization, evidence-grounded explanation, code language models, trustworthy AI}
}

\definecolor{lightblue}{RGB}{235,244,252}
\definecolor{lightgray}{RGB}{245,245,245}
\definecolor{darkred}{RGB}{145,35,35}
\newcommand{\tool}{\textsc{X-LocVul}}
\newcommand{\RQ}[1]{\textbf{RQ#1}}
\newcommand{\pholm}{p_{\mathrm{Holm}}}
\newcommand{\note}[1]{\fcolorbox{darkred}{lightgray}{\parbox{0.94\linewidth}{#1}}}
\newcolumntype{Y}{>{\raggedright\arraybackslash}X}
\setlist[itemize]{leftmargin=1.35em,itemsep=2pt,topsep=4pt}
\setlist[enumerate]{leftmargin=1.5em,itemsep=2pt,topsep=4pt}
\title{\textbf{X-LocVul: An Evidence-Propagation Audit of Cascaded C/C++ Vulnerability Detection, Localization, and Explanation}}
\author{Hoang-Phuong-Linh Le \qquad Quyet-Thang Huynh\\[1mm]
\small School of Information and Communication Technology\\
\small Hanoi University of Science and Technology, Hanoi, Vietnam\\[1mm]
\small \href{mailto:Linh.LHP236037@sis.hust.edu.vn}{\texttt{Linh.LHP236037@sis.hust.edu.vn}}
\quad
\href{mailto:thang.huynhquyet@hust.edu.vn}{\texttt{thang.huynhquyet@hust.edu.vn}}}
\date{September 2026}

\begin{document}
\maketitle
\vspace{-1.5em}

\begin{abstract}
Developer-facing vulnerability analysis must connect a function-level warning
to a source location and a technically supported explanation.  A cascaded
system can, however, propagate weak or incorrect evidence while each component
still appears plausible in isolation.  We present \tool, a measurement-first
audit of a modular C/C++ pipeline comprising a UniXCoder detector, CodeT5 line
generation with source-constrained projection, and a compact Qwen2.5-Coder
explainer.  The study asks whether information passed across the detector--
localizer--explainer interfaces is semantically useful, merely source-resident,
or actively misleading.

Under protocol-level controls intended to reduce duplication and temporal
leakage, UniXCoder has higher point estimates than CodeBERT, but true-CWE
multi-task supervision is statistically indistinguishable from a shuffled-CWE
negative control on PR-AUC, MCC, and vulnerable-class F1.  Raw generated
statements fail normalized exact source matching in 90.83\% of cases.  Fuzzy
projection reaches 83.79\% Top-1 accuracy and exceeds semantic projection by
1.59 percentage points, although its advantage over the retained raw-ranking
baseline is not established.  In a randomized, condition-label-masked,
two-rater evaluation of 50 paired cases, a predicted line improves evidence
support, but adding a predicted CWE does not improve any endpoint.  Oracle
line-and-CWE evidence improves evidence support
($\pholm=0.00453$) and CWE consistency
($\pholm=3.49\times10^{-7}$), while the root-cause difference does not survive
the joint 15-test Holm correction ($\pholm=0.250$).  Unsupported claims remain
common even with oracle evidence.  A chronological stress test retains ranking
signal (ROC-AUC 0.782; PR-AUC 0.110) while the transferred operating threshold
produces zero recall and F1.

The contribution is therefore not a state-of-the-art component claim.  It is
an auditable account of evidence construction, transfer, and consumption that
separates source residency from localization correctness, auxiliary labels
from semantic supervision, and predicted evidence from useful evidence.  The
preprint also exposes reproducibility boundaries: E4 is regenerated from
analysis-complete pre-adjudication score inputs, whereas the E1/E5 sample
manifest, the final Stage-3 prompt/rater
handout, and the Stage-3 profiler log were not recovered.
\end{abstract}


\section{Introduction}

A vulnerability detector that emits only a function-level label leaves most
of the analyst's work unresolved.  The analyst must still identify the
security-relevant statement, understand the unsafe behavior, judge whether the
explanation is supported by the code, and decide whether the proposed repair
addresses the cause rather than a symptom.  Detection, localization, and
explanation answer different questions, yet practical pipelines often compose
them as if a plausible output from one stage were automatically valid evidence
for the next.

This assumption is especially risky for modern code models.  Dataset
duplication can inflate detector performance; a sequence generator can produce
a fluent line that is absent from the source; a nearest-line projection can
return a real but incorrect line; and an explanation model can convert an
incorrect line or coarse weakness label into a confident narrative.  Error is
therefore not merely accumulated.  It can be \emph{transformed}: a detector's
uncertainty becomes a localization request, a generated statement becomes a
source-resident line, and the line becomes an apparently justified security
claim.

\tool{} was initially motivated by LocVul's coupling of function filtering and
sequence-to-sequence localization~\cite{ref1}.  It adds a UniXCoder-based
multi-task detector with a Common Weakness Enumeration (CWE) head and a compact
Qwen2.5-Coder explanation stage.  The architectural composition is useful, but
it is not by itself the scientific novelty: all three model families already
exist.  The central contribution of this preprint is instead an
\emph{evidence-propagation audit}.  We evaluate whether each interface produces
evidence that is present, source-resident, correct, semantically informative,
and useful to the downstream consumer.

This framing leads to four research questions:

\begin{description}[leftmargin=2.7em,style=nextline]
\item[\RQ{1}: Detection and supervision.] Under a common evaluation protocol,
does multi-task CWE supervision change function-level detection, and can any
change be attributed to CWE semantics rather than generic auxiliary-loss
regularization?
\item[\RQ{2}: Localization and grounding.] How often does line generation leave
the source, and do lexical or semantic source projections improve localization
quality rather than merely force source residency?
\item[\RQ{3}: Explanation and evidence use.] Do a predicted line and a predicted
CWE improve human-rated explanation quality, and what does an oracle-evidence
contrast reveal about upstream versus downstream failure?
\item[\RQ{4}: Operational robustness.] What happens when the detector's
operating point is transferred chronologically, and what workload-specific cost
is avoided by filtering before localization and explanation?
\end{description}

The experiments deliberately include controls capable of falsifying an
optimistic account.  Shuffled CWE labels test whether apparent multi-task gains
depend on label meaning.  Exact, fuzzy, and semantic projection distinguish
source membership from correctness.  Four explanation conditions separate
function context, predicted line evidence, predicted CWE anchoring, and oracle
evidence.  A chronological stress test separates ranking quality from the
chosen decision threshold.

The resulting story is primarily a set of disciplined negative findings.
True-CWE supervision does not outperform shuffled-CWE supervision on the
primary endpoints.  Source projection repairs residency but does not establish
semantic correctness.  Predicted CWE anchoring does not improve the explainer.
Oracle evidence helps selected dimensions but does not eliminate unsupported
claims.  A detector can retain non-trivial ranking signal while failing
completely at its transferred operating point.  These findings matter because
they identify which seemingly reasonable claims are \emph{not} supported and
where stronger controls, calibration, or abstention would be required.

This full preprint extends the conference-length version in five ways:

\begin{itemize}
\item it formalizes evidence construction, transfer, and consumption at the
three pipeline interfaces;
\item it reports the complete E1--E6 design, including controls and results
previously compressed into supplementary tables;
\item it publishes all 15 planned E4 tests under one Holm family and regenerates
them from two analysis-complete pre-adjudication rater score files;
\item it distinguishes inferential results from descriptive aggregate records
and historical illustrations; and
\item it gives an explicit reproducibility ledger, including missing artifacts
that prevent full independent reconstruction of E1, E3, E5, and E6.
\end{itemize}

\section{Conceptual Framework: From Stage Outputs to Evidence}

\subsection{Pipeline objects}

Let a C/C++ function be $f=(\ell_1,\ldots,\ell_m)$, where each $\ell_i$ is a
source line.  The three stages produce:

\begin{align}
D(f) &= (p_{\mathrm{vul}},\,\mathbf{p}_{\mathrm{cwe}}), \\
G(f) &= \hat{g}, \\
P(f,\hat{g}) &= (\ell^*,s_{\mathrm{loc}}), \\
X(f,\ell,c) &= x,
\end{align}

where $D$ is the detector, $G$ generates a candidate vulnerable statement,
$P$ projects it onto an actual line with similarity score $s_{\mathrm{loc}}$,
and $X$ generates an explanation $x$ given a function, a line $\ell$, and an
optional CWE anchor $c$.  A function proceeds beyond Stage~1 only if
$p_{\mathrm{vul}}\geq\tau$.

An interface payload is represented as

\begin{equation}
e=(o,\pi,q),
\end{equation}

where $o$ is the output object, $\pi$ is its provenance, and $q$ is any
quality or confidence information available to the consumer.  In the current
prototype, provenance and quality are incomplete: the downstream stage mostly
receives the output object itself.  The audit therefore evaluates properties
that should not be conflated:

\begin{itemize}
\item \textbf{presence}: an output exists;
\item \textbf{residency}: a localized line occurs in the source;
\item \textbf{correctness}: the output matches the annotated vulnerability;
\item \textbf{semantic informativeness}: the information carries meaning beyond
generic regularization or formatting;
\item \textbf{downstream utility}: providing the information improves a
pre-specified endpoint for the next stage; and
\item \textbf{entitlement}: the evidence is sufficient to justify the claim the
system communicates to an analyst.
\end{itemize}

These properties form no automatic implication chain.  In particular,

\begin{equation}
\mathrm{resident}(\ell^*,f)\centernot\Rightarrow
\mathrm{correct}(\ell^*),
\end{equation}

and

\begin{equation}
c\ \mathrm{provided\ to}\ X\centernot\Rightarrow
\mathrm{utility}(c;X)>0.
\end{equation}

Likewise, an auxiliary label can influence optimization without its semantics
being responsible for the change.  This motivates the shuffled-label control
in E1.

\subsection{Three failure loci}

The framework separates three loci that ordinary end-to-end metrics collapse:

\begin{enumerate}
\item \textbf{Evidence construction failure}: the producing stage creates an
incorrect, uncalibrated, or non-source-resident object.
\item \textbf{Evidence transfer failure}: the interface discards uncertainty,
provenance, alternatives, or a reason to abstain, allowing a downstream stage
to treat a tentative output as fact.
\item \textbf{Evidence consumption failure}: the downstream stage ignores,
misinterprets, or overextends otherwise useful evidence.
\end{enumerate}

Oracle contrasts are crucial because they help separate the first two loci
from the third.  If the downstream model improves with oracle evidence but not
predicted evidence, the result is inconsistent with the blanket claim that the
model cannot use evidence.  It instead points to evidence quality, transfer,
or the interaction between predicted inputs and the consumer.

\subsection{Conditional reliability of a cascade}

Let $S_1,S_2,S_3$ denote stage-specific success events under task-appropriate
criteria.  End-to-end success obeys the chain rule

\begin{equation}
\Pr(S_1\cap S_2\cap S_3)
=\Pr(S_1)\Pr(S_2\mid S_1)\Pr(S_3\mid S_1,S_2).
\label{eq:chain}
\end{equation}

Equation~\eqref{eq:chain} makes no independence assumption.  It shows why
isolated component scores are insufficient: a high $\Pr(S_3)$ under oracle
inputs does not imply a high $\Pr(S_3\mid S_1,S_2)$ under predicted inputs,
and a ranking-capable detector does not imply useful $S_1$ at a transferred
threshold.  The study does not estimate a single end-to-end probability because
the retained experiments use different datasets and evaluation units.  Instead,
it estimates targeted interface effects and reports the limitation explicitly.

\section{Related Work}

\subsection{Function-level detection and dataset realism}

Big-Vul~\cite{ref6} and DiverseVul~\cite{ref15} derive C/C++ vulnerability
examples from real fixing commits.  Graph-based systems such as Devign use
explicit program-semantic representations~\cite{ref14}, while CodeBERT and
UniXCoder provide pre-trained code representations~\cite{ref13,ref3}.  Stronger
representations do not remove benchmark bias.  PrimeVul demonstrates that
deduplication, vulnerable/patched pairing, and chronological evaluation can
substantially reduce performance compared with conventional random splits
~\cite{ref2}.  X-LocVul adopts protocol-level grouping and chronological
constraints for the controlled evaluation, while acknowledging that the
sample-level manifest needed to re-audit those assignments was not retained.

Multi-task learning can improve a shared representation through semantic
transfer or through generic regularization~\cite{ref18}.  A single-task ablation
cannot distinguish these explanations.  The shuffled-CWE condition therefore
serves as a negative control: it preserves an auxiliary optimization problem
while destroying the association between a function and its true weakness
category.

\subsection{Line-level localization}

LineVul established transformer-based line-level vulnerability ranking and
developer-inspection metrics~\cite{ref5}; subsequent studies emphasize that
function-level prediction does not automatically imply reliable localization
~\cite{ref16}.  LineVD uses graph neural networks for statement-level detection
~\cite{ref23}.  DeepLineDP addresses line-level defect prediction rather than
security-specific localization~\cite{ref25}, and test-free LLM fault
localization studies general faults~\cite{ref24}.  LocVul uses
sequence-to-sequence generation followed by alignment to the source~\cite{ref1}.
X-LocVul retains that source-projection pattern but audits two distinct
questions: whether generation leaves the source and whether forcing a candidate
back onto the source improves localization correctness.

\subsection{LLM-based vulnerability explanation}

Recent work structures vulnerability reasoning through instruction tuning,
retrieval, program graphs, or reasoning distillation.  Representative examples
include LLMVulExp~\cite{ref8}, Vul-RAG~\cite{ref9}, R2Vul~\cite{ref10},
VulInstruct~\cite{ref11}, LLMxCPG~\cite{ref12}, and VulnLLM-R~\cite{ref22}.
Their different datasets, model sizes, languages, and protocols preclude a
numerical ordering against X-LocVul.  Our contribution is orthogonal: we use a
fixed compact explainer to measure what happens when predicted or oracle
evidence is passed across a cascade interface.

Explanation evaluation should separate fluency from security correctness.
VADER provides a human-evaluated benchmark spanning assessment, detection,
explanation, and remediation~\cite{ref26}.  Work on LLM-as-a-judge methods in
software engineering documents both their potential and their validity risks
~\cite{ref27}.  Empirical guidance calls for reporting model configuration,
prompts, treatments, and human validation~\cite{ref28}.  We therefore use two
human raters and five distinct endpoints; we also disclose that the verbatim
final prompt and original rater handout were not recovered, preventing exact
prompt-level replication.

\section{X-LocVul Architecture}

\begin{figure}[t]
\centering
\includegraphics[width=0.86\textwidth]{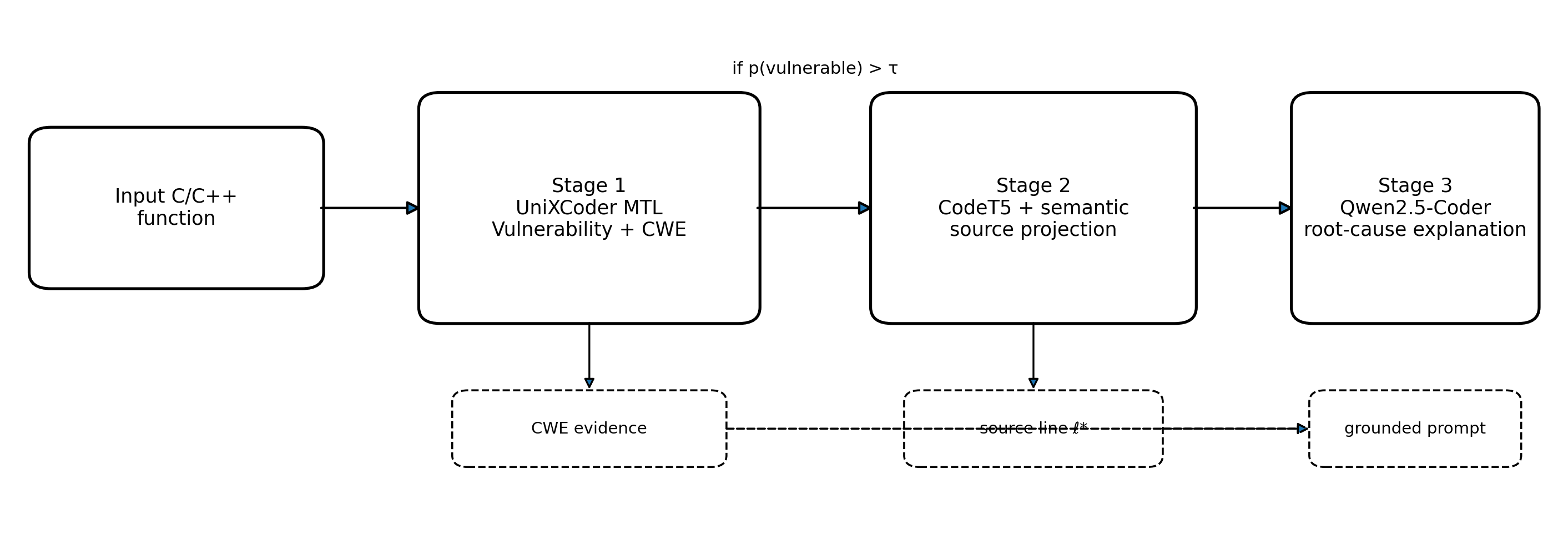}
\caption{The X-LocVul cascade.  Dashed CWE input is an evaluated treatment,
not an assumed benefit.  The audit distinguishes an output's existence,
source residency, correctness, semantic content, and downstream utility.}
\label{fig:architecture}
\end{figure}

\subsection{Stage 1: vulnerability and CWE prediction}

The Stage-1 encoder is initialized from UniXCoder-base.  A shared function
representation $h$ is passed to a binary vulnerability head and a categorical
CWE head.  The joint objective is

\begin{equation}
\mathcal{L}=\lambda_{\mathrm{vul}}\mathcal{L}_{\mathrm{vul}}
+\lambda_{\mathrm{cwe}}\mathcal{L}_{\mathrm{cwe}},
\label{eq:loss}
\end{equation}

where $\mathcal{L}_{\mathrm{vul}}$ is binary cross-entropy and
$\mathcal{L}_{\mathrm{cwe}}$ is categorical cross-entropy.  CodeBERT and
single-task UniXCoder provide architecture and auxiliary-task controls.
True-CWE, shuffled-CWE, hierarchical-CWE, and hard-example-mining variants
probe semantic attribution, label granularity, and the precision--recall
operating point.

\subsection{Stage 2: generation and source projection}

CodeT5 generates a candidate statement $\hat g$.  Since generation can change
identifiers, punctuation, or whitespace, $\hat g$ is not exposed directly as
a location.  Projection selects a source line by either normalized lexical
similarity or embedding cosine similarity:

\begin{align}
\ell^*_{\mathrm{fuzzy}}
&=\arg\max_{\ell_i\in f}\operatorname{sim}_{\mathrm{edit}}(N(\hat g),N(\ell_i)),\\
\ell^*_{\mathrm{semantic}}
&=\arg\max_{\ell_i\in f}
\cos\!\left(\Phi(\hat g),\Phi(\ell_i)\right),
\end{align}

where $N$ normalizes a line and $\Phi$ is the retained embedding function.
Projection guarantees source residency by construction but not ground-truth
correctness.  Generated-Line Hallucination Rate (GLHR) measures the proportion
of raw generated statements that do not exactly match any normalized source
line.  Projection Rescue Rate (PRR) records cases recovered after projection;
Top-$k$ and mean reciprocal rank (MRR) measure localization quality.

The retained aggregate files do not preserve the sample-level ranking outputs
or exact implementation of the row labelled ``raw generation.''  Consequently,
descriptive raw-ranking values are reported, but only the stored paired
bootstrap contrasts are interpreted inferentially.  This boundary is more
conservative than the conference-length account.

\subsection{Stage 3: structured explanation}

The prototype uses Qwen2.5-Coder-1.5B-Instruct~\cite{ref7}.  It receives the
function and, depending on condition, a predicted or oracle line and CWE
anchor.  The intended response contains vulnerability type, source evidence,
root cause, and repair direction.  We use \emph{root-cause explanation} rather
than \emph{causal explanation}: the model performs no formal causal inference.

A lightweight consistency check can flag disagreement between the stated
weakness type and the CWE anchor.  This is a contradiction detector, not an
assurance controller.  It does not establish that either the anchor or the
explanation is correct, estimate joint reliability, or decide when to abstain.

\section{Study Design}

\subsection{Experiments and claims}

\begin{table}[t]
\centering
\caption{Experiment-to-claim map.  ``Reproducibility'' describes the retained
artifact, not the original execution environment.}
\label{tab:design}
\footnotesize
\begin{tabularx}{\textwidth}{@{}p{0.9cm}p{2.2cm}Y Y p{2.2cm}@{}}
\toprule
\textbf{Exp.} & \textbf{RQ/interface} & \textbf{Comparison} &
\textbf{Primary interpretation} & \textbf{Retained reproducibility}\\
\midrule
E1 & RQ1 / detector & CodeBERT, single-task UniXCoder, true/shuffled/parent
CWE, hard mining & Architecture, auxiliary-loss semantics, label granularity,
operating-point trade-off & Aggregate tables; no predictions or split manifest\\
E2 & RQ1 / detector & $\lambda_{\mathrm{cwe}}\in\{0,.1,.2,.5,1\}$ &
Vulnerability/CWE task trade-off & Aggregate mean--SD table\\
E3 & RQ2 / localizer & Raw, exact, fuzzy, semantic & Residency versus
localization quality & Aggregate table and stored paired CIs\\
E4 & RQ3 / explainer & Function, +line, +line+CWE, oracle & Marginal utility of
predicted and correct evidence & Public score inputs and executable regeneration\\
E5 & RQ4 / detector & Fixed validation threshold on chronological test &
Ranking transfer versus operating-point transfer & Five-seed aggregate rows\\
E6 & RQ4 / cascade & Always-run versus Stage-1 filtering & Workload-specific
cost reduction & Aggregate record; Stage-3 VRAM unverified\\
\bottomrule
\end{tabularx}
\end{table}

The experimental design is claim-oriented.  E1--E2 evaluate detection and
the meaning of auxiliary supervision.  E3 evaluates line grounding.  E4
evaluates whether the explainer benefits from the information it receives.
E5 and E6 provide operational stress and cost diagnostics.  Because these
experiments do not share a single end-to-end sample set, we do not multiply
their metrics or report an end-to-end accuracy.

\subsection{Data, splitting, and the retained record}

The original prototype used Big-Vul/PrimeVul-derived C/C++ functions.  The
controlled protocol grouped by project and fixing commit, co-grouped or removed
exact duplicates and high-overlap vulnerable/patched pairs, and constrained
training commits to precede validation/test windows, following the motivation
of PrimeVul~\cite{ref2}.  The final-test threshold was selected on validation
data.

The public artifact does not contain the sample-level manifest, commit IDs,
timestamps, function hashes, or vulnerable/patched pair IDs needed to audit
every assignment.  Moreover, a legacy prose/config record gives E1 counts
whose implied class prevalence conflicts with the majority-clean accuracy in
the retained table.  We therefore omit those unresolved counts and use the
precise phrase \emph{protocol-level leakage controls}.  The aggregate E1
results remain useful for comparing variants within the retained record, but a
future journal submission should regenerate them from predictions and a
published manifest.

\subsection{Configurations}

The retained configuration record gives Stage~1 learning rate
$2\times10^{-5}$, batch size 8, three epochs, weight decay 0.01, and maximum
sequence length 512.  Stage~2 uses CodeT5 with learning rate $5\times10^{-5}$,
10 epochs, beam size 4, and padding labels excluded from loss.  Stage~3 uses
Qwen2.5-Coder-1.5B-Instruct, temperature 0.7, and a 512-token generation
budget.  Trainable comparisons use five seeds where aggregate seed-level
results are retained.

These settings are transcribed from the manuscript and workbook rather than
exported executable run configurations.  Exact software versions, the final
Stage-3 prompt string, and the original rater instruction sheet were not
recovered.  The artifact labels these as limitations instead of reconstructing
them as original evidence.

\subsection{Metrics and statistical analysis}

Under class imbalance, accuracy can be dominated by clean functions.  RQ1
therefore emphasizes vulnerable-class F1, Matthews correlation coefficient
(MCC), PR-AUC, recall, and false-positive rate; ROC-AUC, Brier score, and
expected calibration error are secondary diagnostics.  PR curves are more
informative than ROC curves in strongly imbalanced settings~\cite{ref19}, and
calibration matters because Stage~1 controls cascade routing~\cite{ref17}.

For E1.3 versus E1.5, retained hierarchical-bootstrap intervals compare
PR-AUC, MCC, and vulnerable-class F1.  Paired vulnerable/patched evaluation
uses McNemar tests with Holm adjustment.  E3 reports stored paired-bootstrap
confidence intervals.  Because sample-level E1/E3 predictions were not
retained, these analyses cannot be re-executed from the public artifact and are
identified as retained aggregate results.

E4 is fully regenerated.  Fifty base cases each appear under four conditions,
yielding 200 responses per rater.  The two ratings are averaged within each
base-case/condition cell.  The independent unit for inference is the base case
($n=50$), not the 200 responses.  For five endpoints, we conduct three planned
paired contrasts:

\begin{equation}
(A2-A1),\quad(A3-A2),\quad(A4-A2).
\end{equation}

Two-sided Wilcoxon signed-rank tests use zero splitting.  All $3\times5=15$
tests form one Holm family.  Paired case-level percentile-bootstrap 95\%
intervals use 100,000 resamples with seed 20260911.  Quadratic-weighted Cohen's
$\kappa$ is reported for ordinal scores and unweighted $\kappa$ for the binary
unsupported-claim item~\cite{ref20}, following reliability-reporting guidance
~\cite{ref21}.  Agreement is calculated on original pre-adjudication ratings.

\subsection{Human evaluation treatments}

E4 uses 50 vulnerable functions quota-stratified across ten CWE families, five
base cases per family.  Eligibility required a defensible ground-truth CWE, a
ground-truth vulnerable line, and outputs for all four evidence conditions.
This complete-case requirement supports paired analysis but may induce
survivorship bias by excluding cases where a pipeline branch failed to produce
an output.

\begin{table}[t]
\centering
\caption{E4 evidence conditions.}
\label{tab:e4conditions}
\small
\begin{tabularx}{\textwidth}{@{}p{3.0cm}ccccX@{}}
\toprule
& \textbf{Function} & \textbf{Pred. line} & \textbf{Pred. CWE} &
\textbf{Oracle evidence} & \textbf{Role}\\
\midrule
A1 Function only & \checkmark & -- & -- & -- & Baseline context\\
A2 Function + line & \checkmark & \checkmark & -- & -- & Marginal utility of predicted line\\
A3 Function + line + CWE & \checkmark & \checkmark & \checkmark & -- & Marginal utility of predicted CWE\\
A4 Oracle & \checkmark & -- & -- & line + CWE & Upper-bound evidence contrast\\
\bottomrule
\end{tabularx}
\end{table}

Responses were assigned randomized identifiers and presented without condition
labels.  The evaluation is therefore described as randomized and
condition-label-masked, not fully blinded: raters could often infer whether a
line or CWE had been supplied from the response context.  Two raters scored
root-cause correctness, evidence support, CWE consistency, and repair
usefulness on 1--5 ordinal scales and unsupported claims as a binary outcome.

\section{Results}

\subsection{RQ1: detection, semantic supervision, and operating points}

\begin{table}[p]
\centering
\caption{Controlled E1 comparison.  Trainable rows are mean $\pm$ SD over five
seeds where retained; E1.3/E1.5 pooled primary metrics include hierarchical-
bootstrap 95\% intervals.  Values are retained aggregates.}
\label{tab:e1}
\footnotesize
\begin{tabularx}{\textwidth}{@{}lYrrrrr@{}}
\toprule
\textbf{Run} & \textbf{Configuration} & \textbf{MCC} & \textbf{Vul. F1 (\%)} &
\textbf{PR-AUC} & \textbf{Recall (\%)} & \textbf{FPR (\%)}\\
\midrule
E1.0a & Majority-clean & 0.000 & 0.00 & 0.000 & 0.00 & 0.00\\
E1.0b & Lexical-logistic & 0.129 & 10.15 & 0.089 & 53.33 & 16.96\\
E1.1 & CodeBERT, single-task & $0.071\pm0.066$ & $4.10\pm4.29$ & $0.085\pm0.041$ & $2.27\pm2.43$ & $0.06\pm0.06$\\
E1.2 & UniXCoder, single-task & $0.115\pm0.061$ & $7.33\pm4.77$ & $0.134\pm0.003$ & $4.20\pm2.88$ & $0.11\pm0.09$\\
E1.3 & UniXCoder + true CWE & 0.145 & 7.73 & 0.146 & $4.69\pm2.67$ & $0.11\pm0.06$\\
E1.5 & UniXCoder + shuffled CWE & 0.154 & 8.20 & 0.147 & $4.74\pm0.65$ & $0.09\pm0.02$\\
E1.6 & UniXCoder + parent CWE & $0.075\pm0.065$ & $3.84\pm3.75$ & $0.110\pm0.040$ & $2.07\pm2.06$ & $0.04\pm0.04$\\
E1.4 & True CWE + hard mining & $0.129\pm0.030$ & $13.37\pm3.23$ & $0.103\pm0.007$ & $13.04\pm7.97$ & $1.31\pm1.12$\\
\bottomrule
\end{tabularx}
\end{table}

\subsubsection{Architecture point estimates are not semantic evidence}

Within the retained protocol, UniXCoder variants have higher point estimates
than CodeBERT.  Single-task UniXCoder raises MCC from 0.071 to 0.115 and PR-AUC
from 0.085 to 0.134.  Adding true-CWE supervision raises the corresponding
point estimates to 0.145 and 0.146.  We do not label the encoder difference
significant because a direct paired interval is unavailable.

More importantly, E1.5 defeats the simple interpretation that CWE semantics
caused the multi-task change.  Shuffled-CWE supervision matches or exceeds
true-CWE supervision on MCC, vulnerable-class F1, and PR-AUC.  The direct
E1.3--E1.5 intervals in \cref{tab:true-shuffled} include zero for every primary
endpoint.  The supported conclusion is therefore narrow: an auxiliary loss
changes optimization, but these data do not attribute the change to correct
CWE meaning.

\begin{table}[t]
\centering
\caption{True-CWE minus shuffled-CWE hierarchical-bootstrap contrasts.}
\label{tab:true-shuffled}
\begin{tabular}{@{}lrrr@{}}
\toprule
\textbf{Endpoint} & \textbf{E1.3 true} & \textbf{E1.5 shuffled} &
\textbf{Difference [95\% CI]}\\
\midrule
PR-AUC & 0.1464 & 0.1467 & $-0.0002\;[-0.0082,0.0075]$\\
MCC & 0.1445 & 0.1535 & $-0.0090\;[-0.0408,0.0248]$\\
Vulnerable F1 & 0.0773 & 0.0820 & $-0.0047\;[-0.0242,0.0148]$\\
\bottomrule
\end{tabular}
\end{table}

\subsubsection{Granularity and hard-mining sensitivity}

The parent-CWE variant has lower MCC, F1, and PR-AUC than single-task
UniXCoder.  This result demonstrates sensitivity to the chosen label mapping;
it does not prove that coarse supervision is intrinsically harmful.  Class
imbalance, mapping ambiguity, loss weighting, and the number of auxiliary
classes are alternative explanations.

Hard-example mining produces the highest F1 (13.37\%) by raising recall to
13.04\%, but FPR increases approximately twelve-fold relative to E1.3 and seed
variance is much larger.  The result is an operating-point trade-off, not an
unqualified improvement.  In a security workflow, whether this trade is
desirable depends on the cost of missed vulnerabilities, analyst capacity,
and the expense of downstream generation.

\subsubsection{CWE-loss sweep}

\begin{table}[t]
\centering
\caption{E2 CWE-loss sweep (mean $\pm$ SD over five retained runs).}
\label{tab:e2}
\small
\begin{tabular}{@{}crrrrr@{}}
\toprule
$\lambda_{\mathrm{cwe}}$ & \textbf{Vul. F1} & \textbf{CWE macro-F1} &
\textbf{MCC} & \textbf{PR-AUC} & \textbf{ECE}\\
\midrule
0.0 & $.0844\pm.0506$ & $.0206\pm.0045$ & $.1294\pm.0732$ & $.1171\pm.0424$ & $.0788\pm.1509$\\
0.1 & $.0740\pm.0370$ & $.0340\pm.0047$ & $.1297\pm.0375$ & $.1334\pm.0067$ & $.0112\pm.0006$\\
0.2 & $.0749\pm.0467$ & $.0351\pm.0072$ & $.1350\pm.0504$ & $.1328\pm.0088$ & $.0109\pm.0005$\\
0.5 & $.0543\pm.0248$ & $.0415\pm.0080$ & $.1089\pm.0345$ & $.1303\pm.0048$ & $.0108\pm.0006$\\
1.0 & $.0701\pm.0073$ & $.1011\pm.0306$ & $.1389\pm.0148$ & $.1344\pm.0103$ & $.0107\pm.0008$\\
\bottomrule
\end{tabular}
\end{table}

Increasing $\lambda_{\mathrm{cwe}}$ from 0 to 1 raises CWE macro-F1 from
2.06\% to 10.11\%, but vulnerability F1 is non-monotonic and ends below its
$\lambda=0$ value.  PR-AUC varies little among non-zero weights.  E2 therefore
supports a task trade-off, not a monotonic transfer benefit.

\subsubsection{Cross-project and vulnerable/patched diagnostics}

On disjoint BigVul projects, UniXCoder PR-AUC can be higher than in the primary
test (for example, E1.2: 0.472 versus 0.134), while recall at the transferred
threshold is near zero for most configurations.  The ranking/decision mismatch
is consistent with threshold and calibration transfer being major contributors,
although representation shift cannot be excluded.

The paired vulnerable/patched diagnostic is more concerning.  Paired accuracy
is approximately chance (49.7--50.3\%) across learned configurations, and no
comparison against CodeBERT survives the retained Holm-adjusted McNemar tests.
Current models therefore do not reliably distinguish a vulnerable function
from its patched counterpart under this diagnostic.  This finding limits claims
that the classifiers have learned the semantic edit that removes the
vulnerability.

\paragraph{Answer to RQ1.}
The retained results do not support a claim that correct CWE supervision
improves vulnerability detection.  Multi-task changes are non-monotonic,
indistinguishable from shuffled-label regularization on primary endpoints, and
sensitive to label granularity and operating point.

\subsection{RQ2: generated-line residency and localization}

\begin{table}[t]
\centering
\caption{E3 localization strategies.  Brackets are retained 95\% intervals.}
\label{tab:e3}
\footnotesize
\resizebox{\textwidth}{!}{%
\begin{tabular}{@{}lccccc@{}}
\toprule
\textbf{Strategy} & \textbf{GLHR} & \textbf{PRR} & \textbf{Top-1} &
\textbf{Top-5} & \textbf{MRR}\\
\midrule
Raw generation & -- & -- & $.8306\;[.8053,.8559]$ & $.8331\;[.8081,.8584]$ & $.8318\;[.8069,.8570]$\\
Normalized exact match & $.9083\;[.9008,.9158]$ & -- & $.4160\;[.3817,.4470]$ & $.4173\;[.3830,.4480]$ & $.4166\;[.3825,.4475]$\\
Fuzzy projection & -- & $.0651\;[.0608,.0693]$ & $.8379\;[.8127,.8630]$ & $.8580\;[.8353,.8817]$ & $.8465\;[.8225,.8711]$\\
Semantic projection & -- & $.0657\;[.0611,.0704]$ & $.8221\;[.7957,.8487]$ & $.8490\;[.8259,.8739]$ & $.8331\;[.8097,.8591]$\\
\bottomrule
\end{tabular}}
\end{table}

Normalized exact matching fails for 90.83\% of raw generations, showing that
the generator frequently produces a statement that is not literally present
in the function.  Exact-match localization accordingly reaches only 41.60\%
Top-1.  Fuzzy and semantic projections always return source lines and recover
much higher ranking metrics.

The central comparison is not ``projected versus hallucinated,'' because
source residency alone is insufficient.  Fuzzy projection exceeds semantic
projection by 1.59 percentage points in Top-1 (stored paired-bootstrap 95\% CI
$[0.23,2.94]$ for fuzzy minus semantic) and 1.34 points in MRR
($[0.16,2.54]$).  Fuzzy versus raw ranking is inconclusive for Top-1
(0.73 points, CI $[-0.76,2.31]$) and nearly touches zero for MRR (1.47 points,
CI $[-0.03,3.03]$).  Fuzzy is therefore preferred to semantic projection for
simplicity and the observed paired advantage, but a demonstrated improvement
over the retained raw-ranking baseline is not claimed.

\paragraph{Answer to RQ2.}
Generation frequently leaves the source.  Projection restores source residency,
and lexical projection outperforms semantic projection in the retained paired
contrasts.  The experiment does not establish that projection beats raw
ranking, and source residency remains distinct from correct localization.

\subsection{RQ3: explanation grounding}

\begin{table}[t]
\centering
\caption{E4 two-rater condition means over 50 paired base cases.  The first
four endpoints use 1--5 scales; unsupported claim is a binary rate.}
\label{tab:e4means}
\begin{tabular}{@{}lccccc@{}}
\toprule
\textbf{Condition} & \textbf{Root cause} & \textbf{Evidence} & \textbf{CWE} &
\textbf{Repair} & \textbf{Unsupported}\\
\midrule
A1 Function only & 1.40 & 2.27 & 1.40 & 1.33 & 0.90\\
A2 Function + line & 1.34 & 3.42 & 1.72 & 1.33 & 0.90\\
A3 Function + line + CWE & 1.40 & 3.51 & 1.40 & 1.27 & 0.86\\
A4 Oracle evidence & 1.86 & 4.43 & 3.70 & 1.59 & 0.87\\
\bottomrule
\end{tabular}
\end{table}

\Cref{tab:e4means} shows three distinct patterns.  First, a predicted line
raises evidence support from 2.27 to 3.42 but does not improve root-cause or
repair means.  Second, adding predicted CWE to the line does not improve the
condition means and reduces CWE consistency from 1.72 to 1.40.  Third, oracle
evidence produces much higher evidence and CWE scores, yet root-cause and repair
scores remain low and the unsupported-claim rate remains 0.87.

\begin{longtable}{@{}p{3.2cm}p{2.7cm}rrr@{}}
\caption{All 15 pre-specified E4 contrasts regenerated from public rater-score inputs.
Intervals are paired case-level percentile-bootstrap 95\% CIs; Holm adjustment
uses one 15-test family.}\label{tab:e4all}\\
\toprule
\textbf{Contrast} & \textbf{Endpoint} & \textbf{Mean diff.} & \textbf{95\% CI} &
$\boldsymbol{\pholm}$\\
\midrule
\endfirsthead
\toprule
\textbf{Contrast} & \textbf{Endpoint} & \textbf{Mean diff.} & \textbf{95\% CI} &
$\boldsymbol{\pholm}$\\
\midrule
\endhead
A2--A1 & Root cause & $-0.06$ & $[-0.42,0.31]$ & 1.000\\
A2--A1 & Evidence & $1.15$ & $[0.62,1.67]$ & \textbf{0.004530}\\
A2--A1 & CWE & $0.32$ & $[0.00,0.68]$ & 1.000\\
A2--A1 & Repair & $0.00$ & $[-0.35,0.38]$ & 1.000\\
A2--A1 & Unsupported & $0.00$ & $[-0.08,0.08]$ & 1.000\\
\midrule
A3--A2 & Root cause & $0.06$ & $[-0.24,0.35]$ & 1.000\\
A3--A2 & Evidence & $0.09$ & $[-0.25,0.43]$ & 1.000\\
A3--A2 & CWE & $-0.32$ & $[-0.70,0.05]$ & 1.000\\
A3--A2 & Repair & $-0.06$ & $[-0.33,0.21]$ & 1.000\\
A3--A2 & Unsupported & $-0.04$ & $[-0.13,0.04]$ & 1.000\\
\midrule
A4--A2 & Root cause & $0.52$ & $[0.16,0.91]$ & 0.250386\\
A4--A2 & Evidence & $1.01$ & $[0.56,1.47]$ & \textbf{0.004530}\\
A4--A2 & CWE & $1.98$ & $[1.52,2.42]$ & \textbf{$3.4927\times10^{-7}$}\\
A4--A2 & Repair & $0.26$ & $[-0.03,0.58]$ & 1.000\\
A4--A2 & Unsupported & $-0.03$ & $[-0.14,0.07]$ & 1.000\\
\bottomrule
\end{longtable}

Only three of the 15 adjusted tests are supported at 0.05.  A2--A1 establishes
that the predicted line improves evidence support.  A4--A2 establishes that
oracle line-and-CWE evidence improves evidence support and CWE consistency.
The positive root-cause mean difference under oracle evidence has an interval
excluding zero, but its Wilcoxon result does not survive the joint family
correction; it must not be reported as statistically supported.  No contrast
reduces unsupported claims.

Before adjudication, exact agreement for root cause, evidence, CWE, repair, and
unsupported claims is 88.0\%, 82.0\%, 89.0\%, 88.5\%, and 88.5\%.  Quadratic-
weighted $\kappa$ for the ordinal endpoints is 0.919, 0.877, 0.901, and 0.781;
unweighted $\kappa$ for unsupported claims is 0.446.  The latter combination
of high exact agreement and moderate $\kappa$ is consistent with the strong
prevalence of unsupported-claim judgments.

\paragraph{Answer to RQ3.}
Predicted CWE anchoring provides no supported improvement beyond predicted-line
input.  Oracle evidence improves evidence support and CWE consistency, locating
an upstream evidence-quality opportunity.  The persistent unsupported-claim
rate shows that better upstream evidence is not sufficient for safe explanation
generation; downstream claim control or abstention is also required.

\subsection{RQ4: chronological threshold transfer and cascade cost}

\subsubsection{Chronological stress test (E5)}

\begin{table}[H]
\centering
\caption{E5 chronological/de-duplicated stress test at the fixed validation
threshold (five-seed mean).  Precision is recorded as zero under the analysis
pipeline's zero-division rule when no true-positive operating output is
obtained.}
\label{tab:e5}
\begin{tabular}{@{}ccccccc@{}}
\toprule
\textbf{Accuracy} & \textbf{MCC} & \textbf{Precision} & \textbf{Recall} &
\textbf{F1} & \textbf{PR-AUC} & \textbf{ROC-AUC}\\
\midrule
0.9655 & $-0.0018$ & 0.000 & 0.000 & 0.000 & 0.1097 & 0.7821\\
\bottomrule
\end{tabular}
\end{table}

E5 is a stress test of operating-point transfer, not a retrained chronological
benchmark.  ROC-AUC remains 0.782 and PR-AUC 0.110 against 3.43\% prevalence,
but the fixed threshold yields zero vulnerable recall and F1 across all five
seeds.  Mean FPR is 0.000195, showing an extremely conservative transferred
operating point.  The model retains ranking information that the fixed
threshold fails to convert into actionable detections.  At the same time, the
PR-AUC decrease relative to E1.3 is consistent with temporal ranking
degradation, so threshold transfer cannot be declared the sole cause.

Sample-level probabilities and a held-out chronological calibration window are
not in the retained artifact.  We therefore do not present a post-hoc optimal
threshold, which would be both irreproducible and optimistically biased.

\subsubsection{Runtime and workload filtering (E6)}

\begin{table}[H]
\centering
\caption{Retained E6 aggregate runtime records on an RTX 5090.  Stage-3 peak
VRAM is historical reported-only and not raw-profiler verified.}
\label{tab:e6}
\begin{tabular}{@{}lccc@{}}
\toprule
\textbf{Metric} & \textbf{Stage 1} & \textbf{Stage 2} & \textbf{Stage 3}\\
\midrule
Model & UniXCoder & CodeT5 & Qwen2.5-Coder-1.5B\\
Median latency (ms) & 2.65 & 282.84 & 502.92\\
Throughput (functions/s) & 377.89 & 3.54 & 1.99\\
Peak VRAM (MB) & 490.27 & 914.05 & 3891.06$^{*}$\\
\bottomrule
\multicolumn{4}{l}{\footnotesize $^{*}$Prior manuscript value; original profiler log unavailable.}
\end{tabular}
\end{table}

In a retained balanced workload of 200 functions (100 clean, 100 vulnerable),
Stage~1 filters 95 clean functions, so 105 proceed to the two generative stages.
The aggregate summary reports average latency falling from 788.41 to
415.17~ms per input, a 47.3\% workload-specific reduction.  This is arithmetic
and workload evidence, not a general production-cost estimate: the benefit
depends on prevalence, threshold, batching, hardware, queueing, and the cost of
false negatives.  E5 further shows that aggressive filtering can erase
vulnerable recall under distribution shift.

\paragraph{Answer to RQ4.}
The fixed threshold is not temporally robust despite retained ranking signal.
Filtering can nearly halve compute on the balanced retained workload, but the
cost claim and the detection-risk claim must be evaluated jointly.

\FloatBarrier

\section{Cross-Stage Synthesis}

\subsection{A claim--evidence ledger}

The four RQs can be summarized as a ledger rather than a performance ranking.
For every attractive mechanism claim, the study asks what observation would
entitle it and whether that observation is present.

\begin{table}[!htbp]
\centering
\caption{Claim ledger.  ``Not supported'' does not mean the mechanism can never
work; it means the present evidence does not entitle the stated claim.}
\label{tab:ledger}
\small
\begin{tabularx}{\textwidth}{@{}Y Y Y >{\raggedright\arraybackslash}p{2.3cm}@{}}
\toprule
\textbf{Candidate claim} & \textbf{Required evidence} & \textbf{Observed audit result} &
\textbf{Status}\\
\midrule
Correct CWE supervision improves detection & True CWE exceeds both single-task
and shuffled-CWE controls & True CWE does not exceed shuffled CWE on primary
paired intervals & Not supported\\
Semantic projection is superior to lexical projection & Higher paired Top-$k$/MRR
under common candidates & Semantic is lower by 1.59 Top-1 and 1.34 MRR points &
Contradicted here\\
Projection improves localization over raw ranking & Paired interval excluding
zero & Fuzzy--raw intervals include/touch zero & Not supported\\
Predicted CWE improves explanation & A3--A2 improves pre-specified endpoints &
No endpoint survives Holm; CWE mean decreases & Not supported\\
Correct evidence is useful & A4--A2 improves evidence-sensitive endpoints &
Evidence support and CWE consistency improve & Supported for two endpoints\\
Correct evidence makes explanations safe & Oracle reduces unsupported claims &
Unsupported rate remains 0.87; adjusted contrast is 1.0 & Not supported\\
Ranking quality implies a usable deployed decision & Fixed threshold preserves
recall/F1 under time shift & ROC-AUC 0.782 but recall/F1 0 & Contradicted here\\
\bottomrule
\end{tabularx}
\end{table}

This ledger is the main scientific identity of X-LocVul.  The paper does not
celebrate a cascade merely because its modules can be connected.  It identifies
the evidence needed for each cross-stage claim and reports when the required
contrast fails.

\FloatBarrier

\subsection{Why the negative controls matter}

Without E1.5, the change from single-task UniXCoder to true-CWE multi-task
learning could be described as a semantic benefit.  The shuffled control shows
that this attribution is not earned.  Without the exact-match diagnostic,
source projection could hide the fact that the generator frequently leaves the
source.  Without A4, the absence of A3 benefit could be blamed entirely on the
explainer's inability to use structured evidence.  Oracle evidence rejects
that blanket explanation for evidence support and CWE consistency while
revealing that unsupported claims persist downstream.

Negative controls do more than make conclusions conservative.  They localize
mechanisms.  E1.5 targets the semantics of the auxiliary signal; A3--A2 targets
the incremental value of predicted CWE; A4--A2 targets the joint quality ceiling
of line and CWE evidence; E5 targets operating-point transfer.  Their combined
value is higher than another uncontrolled component comparison would be.

\subsection{Failure propagation patterns}

The experiments reveal four recurring patterns:

\begin{enumerate}
\item \textbf{Regularization masquerading as semantics.}  An auxiliary task can
move metrics even when its labels are shuffled.  Semantic claims therefore need
meaning-destroying controls.
\item \textbf{Grounding masquerading as correctness.}  Projection guarantees a
line is in the function, which is valuable for inspectability, but does not
guarantee it is the vulnerable line.
\item \textbf{Availability masquerading as utility.}  A predicted CWE can be
present in the prompt without measurably improving explanation quality.
\item \textbf{Ranking masquerading as a deployable decision.}  ROC-AUC and
PR-AUC describe ordering across thresholds; a deployed system acts at one
operating point, which can fail under shift.
\end{enumerate}

These are not peculiarities of one vulnerability pipeline.  They are general
risks in cascaded AI systems whose outputs are reinterpreted as evidence by
later components.

\section{Design Implications}

\subsection{Evidence should carry provenance and alternatives}

The current interfaces transfer a chosen object---a top CWE or one projected
line---more readily than they transfer uncertainty.  A stronger interface
would pass a structured record such as

\begin{equation}
e_2=\bigl(\{(\ell_i,s_i)\}_{i=1}^{k},\,m,\,a,\,\pi\bigr),
\end{equation}

where the top-$k$ source lines retain scores $s_i$, $m$ is the margin between
leading candidates, $a$ records cross-method agreement, and $\pi$ identifies
the generating model, checkpoint, normalization, and function hash.  The
downstream model could then distinguish a decisive localization from a near
tie, and an auditor could reconstruct why a line was supplied.

Likewise, Stage~1 should pass calibrated class and CWE distributions rather
than only top labels.  A downstream explanation should not state a specific
weakness category more strongly than the available distribution supports.

\subsection{Abstention must be evaluated, not merely added}

The high unsupported-claim rate motivates abstention, but simply inserting a
confidence threshold is not enough.  A future controller should combine at
least detector calibration, localization margin, lexical--semantic agreement,
CWE entropy, and line--explanation consistency.  Evaluation should report:

\begin{itemize}
\item risk--coverage curves for both detection and explanation;
\item false-assurance rate: confident outputs with incorrect or unsupported
evidence;
\item selective calibration under project and temporal shift;
\item end-to-end coverage after each gate; and
\item compute saved versus vulnerabilities missed.
\end{itemize}

Oracle evidence provides an upper-bound diagnostic, not a deployable solution.
The engineering objective is to approach its benefit while preserving honest
uncertainty and abstaining when evidence is insufficient.

\subsection{Threshold governance is part of model governance}

E5 shows why a model artifact cannot be separated from its operating policy.
A threshold selected in one window can become unusable even when ranking
remains non-random.  Deployment should therefore version the calibration data,
threshold, target prevalence, expected false-positive budget, and the date and
conditions under which the operating point was validated.  Recalibration
should be triggered by monitored drift rather than performed silently on the
test set.

\subsection{Recommended next experiment}

The highest-value next study is not another encoder swap.  It is an end-to-end,
sample-aligned evaluation with three additions:

\begin{enumerate}
\item publish predictions, probabilities, top-$k$ lines, prompts, responses,
and provenance under one immutable sample manifest;
\item implement an evidence-aware selective controller and compare fixed
threshold, recalibrated threshold, and abstaining policies under chronological
and cross-project shifts; and
\item evaluate explanation outcomes on all routed inputs, including pipeline
failures, to remove complete-case survivorship bias.
\end{enumerate}

This would turn the current interface audit into an end-to-end assurance study.

\section{Historical Qualitative Illustration}

An earlier prototype log described tcpdump CVE-2017-13036 (CWE-125).  Stage~1
selected the correct vulnerability family; Stage~2 projected to
\texttt{if (ndo->ndo\_vflag > 1) \{}; and Stage~3 described a missing bounds
check but called the resulting out-of-bounds read a ``buffer overflow.''  The
example illustrates that a fluent and partially useful explanation can still
misstate the security taxonomy, motivating the CWE-consistency treatment.

\note{\textbf{Provenance boundary.} This is a non-inferential historical
illustration outside E4.  It was not selected into the 50 E4 base cases, does
not occur in the 200 rater responses, and contributes to no statistic in this
paper.  Its original run log and stable sample identifier were not recovered.}

The example is retained only to explain design motivation.  The statistical
evidence for CWE anchoring comes exclusively from A3--A2 and A4--A2, not from
this case.

\section{Threats to Validity}

\subsection{Construct validity}

Function-level labels do not capture interprocedural exploitability, reachability,
configuration, or environmental constraints.  Vulnerable-class F1 and MCC
measure classification but not analyst utility.  Top-$k$ localization measures
agreement with an annotated line, while some vulnerabilities involve multiple
statements or data-flow paths.  Human-rated explanations are multidimensional;
the 1--5 scales compress technical distinctions and unsupported-claim coding
depends on the evaluation instrument.

GLHR measures exact source membership after normalization, not hallucination in
the broader factual sense.  A generated statement can be absent yet semantically
equivalent to a source line; conversely, a projected source line can be present
but wrong.  We retain the acronym used in the original analysis but interpret it
narrowly as generated-line non-residency.

\subsection{Internal validity}

The E1 variants are described as sharing preprocessing, checkpoint selection,
and threshold procedures, but sample-level predictions and run configurations
are not retained.  The shuffled-label control is strong for semantic
attribution, yet one shuffled realization may not characterize the full
distribution of random auxiliary tasks.  Parent-CWE performance may be affected
by class distribution, mapping rules, or loss weighting rather than granularity
alone.

E4 uses one generated response per condition.  Stochastic variation from
temperature 0.7 is therefore not estimated.  Raters saw randomized,
condition-label-masked responses, but content could reveal the treatment.  The
original handout is unavailable, so exact replication of their interpretation
of the rubric is impossible.  Averaging two ordinal ratings is a pragmatic
condition summary; paired Wilcoxon inference is performed at the base-case
level and does not turn that average into an interval-scale claim.

The complete-output eligibility rule for E4 can create survivorship bias.  A
system that fails to generate under difficult conditions may be underrepresented.
Future studies should code missing outputs as operational outcomes rather than
exclude them.

\subsection{Conclusion validity}

The E4 analysis pre-specifies 15 comparisons and controls them jointly with
Holm.  Reporting all 15 avoids outcome selection.  Confidence intervals and
Wilcoxon tests summarize different aspects of paired differences; as the
A4--A2 root-cause result shows, an interval may exclude zero while the planned
rank-based test is not significant after family correction.  Our inferential
claim follows the corrected test.

E1 and E3 confidence intervals are retained aggregate outputs.  Without
sample-level predictions, their resampling units and code cannot be independently
audited.  They are consequently treated more cautiously than E4.  Small
point-estimate differences must not be converted into claims of superiority.

\subsection{External validity}

The study concerns C/C++ functions and selected CWE families.  It excludes
interprocedural vulnerabilities, configuration errors, memory-safe languages,
repository-level context, and dynamic exploit evidence.  Qwen2.5-Coder-1.5B is
a compact explainer; larger, newer, or tool-using models may respond differently.
The runtime workload is balanced and measured on one GPU family, so its 47.3\%
latency reduction is not a production-general estimate.

No common-protocol head-to-head evaluation against every recent detector,
localizer, and explainer is available.  The study supports claims about controls
and interfaces within X-LocVul, not state-of-the-art ranking across heterogeneous
systems.

\subsection{Reproducibility validity}

The public artifact has asymmetric strength.  E4 includes analysis-complete
pre-adjudication score inputs, checksums, deterministic regeneration, all 15
tests, and derived tables.  The internally retained full exports also contain
code snippets, model responses, and free-text notes; those fields are not
redistributed in the public release.  E1,
E2, E3, E5, and E6 are largely aggregate exports.  The E1/E5 sample manifest,
exact Stage-3 prompt, original rater handout, and Stage-3 profiler log are
missing.  The Stage-3 VRAM value is therefore reported-only.  A legacy E1
sample-count record also conflicts with the retained majority baseline; the
unresolved counts are omitted.

These limitations reduce exact reproducibility, but making them explicit
prevents a stronger and false claim that the artifact contains data it does
not.  The compatibility manifest maps every numerical statement to its actual
support level.

\section{Artifact and Evidence Boundaries}

\begin{table}[t]
\centering
\caption{Reproducibility ledger for the arXiv package.}
\label{tab:repro}
\small
\begin{tabularx}{\textwidth}{@{}p{1.1cm}Y Y >{\raggedright\arraybackslash}p{2.5cm}@{}}
\toprule
\textbf{Exp.} & \textbf{Included} & \textbf{Missing} & \textbf{Allowed claim}\\
\midrule
E1 & Canonical aggregate table; true--shuffled intervals; cross-project and
paired summaries & Predictions, confusion matrices, manifest, executable configs &
Within-record comparison only\\
E2 & Five-weight aggregate table & Seed predictions and training logs &
Descriptive trade-off\\
E3 & Aggregate metrics and paired-CI summary & Candidate rankings and
projection code & Descriptive plus stored paired contrasts\\
E4 & Two public rater-score files, hashes, merge, means, CIs, 15 tests, $\kappa$,
script & Full response/code context; final verbatim prompt; original rater handout &
Score-input-regenerated inference\\
E5 & Five seed rows & Probabilities, labels, calibration window, manifest &
Fixed-threshold diagnostic\\
E6 & Aggregate latency/throughput and workload summary & Request-level traces;
Stage-3 profiler log & Workload-specific descriptive cost\\
\bottomrule
\end{tabularx}
\end{table}

The three excluded preliminary \texttt{S1\_*} exports are not part of the public data
package because they belong to a superseded or unresolved protocol and contain
damaged cells.  Their filenames and SHA-256 hashes are retained in a provenance
note, making the exclusion transparent without presenting incompatible results
as if they were part of E1.

No third-party source snippet, generated response, or free-text rater note is
distributed in the public score inputs.  The transformation retains exactly the
case key, condition, and five pre-adjudication ratings required to regenerate
every reported E4 statistic.  Original-export hashes and the field-selection
rule are documented in the package provenance record.

\FloatBarrier

\section{Conclusion}

X-LocVul reframes a cascaded vulnerability pipeline as an evidence-propagation
problem.  The audit finds that true-CWE supervision is indistinguishable from a
shuffled-label control on the primary detection endpoints; generation often
leaves the source; fuzzy projection outperforms semantic projection but does not
establish superiority over retained raw ranking; predicted CWE anchoring does
not improve explanation; and a fixed threshold fails under chronological shift
despite non-trivial ranking metrics.  Oracle evidence improves evidence support
and CWE consistency, yet unsupported claims remain frequent.

The broader lesson is methodological.  Source residency is not localization
correctness.  An auxiliary label is not automatically semantic evidence.  A
provided feature is not necessarily useful to its consumer.  Ranking ability
is not a deployed decision.  And correct upstream evidence is not sufficient
for safe downstream claims.  Reliable cascades therefore require explicit
controls, calibrated operating policies, provenance-bearing interfaces, and
evaluated abstention.

This preprint's strongest reproducible result is E4, rebuilt from public
analysis-complete rater-score inputs with a complete 15-test family.  Other
experiments remain bounded by aggregate
records and disclosed missing provenance.  Recovering the unified sample
manifest, predictions, final prompts, and profiler logs is the necessary next
step before making stronger end-to-end or state-of-the-art claims.

\section*{Data and Code Availability}

The reproducibility artifact and data supporting this study---including the
narrative supplement, canonical aggregate tables, analysis-complete E4 score
inputs, deterministic E4 regeneration, checksums, provenance notes, and the
machine-checked paper--supplement mapping---are archived on Zenodo as
\emph{X-LocVul: Reproducibility Artifact and Data for Vulnerability Detection,
Localization, and Explanation}~\cite{ref29}.

\section*{Ethics and Responsible Use}

The work evaluates source-code vulnerability analysis and should not be read as
authorizing automated repair or deployment decisions.  False negatives can
leave exploitable code unreviewed, while unsupported explanations can misdirect
analysts.  Human review, secure handling of vulnerability data, and appropriate
disclosure practices remain necessary.

\section*{Declaration of Competing Interests}

The authors declare no competing interests.

\appendix

\section{Expanded Cross-Project Results}

\begin{table}[h]
\centering
\caption{Retained cross-project BigVul aggregate results at the PrimeVul-selected
threshold.}
\label{tab:crossproject}
\scriptsize
\begin{tabularx}{\textwidth}{@{}lYrrrrr@{}}
\toprule
\textbf{Run} & \textbf{Model/variant} & \textbf{MCC} & \textbf{Recall} &
\textbf{F1} & \textbf{PR-AUC} & \textbf{ROC-AUC}\\
\midrule
E1.0b & Lexical-logistic & .1151 & .3229 & .1755 & .1163 & .6437\\
E1.1 & CodeBERT & $.0058\pm.0050$ & $.0001\pm.0001$ & $.0002\pm.0002$ & $.1179\pm.0286$ & $.6668\pm.0652$\\
E1.2 & UniXCoder single & $.0287\pm.0123$ & $.0022\pm.0015$ & $.0043\pm.0031$ & $.4720\pm.0346$ & $.8965\pm.0138$\\
E1.3 & UniXCoder true CWE & $.0261\pm.0106$ & $.0021\pm.0013$ & $.0041\pm.0025$ & $.4488\pm.0808$ & $.8508\pm.0420$\\
E1.4 & True CWE + hard mining & $.1251\pm.1230$ & $.0684\pm.1004$ & $.1009\pm.1332$ & $.5225\pm.0435$ & $.8811\pm.0297$\\
E1.5 & UniXCoder shuffled CWE & $.0333\pm.0033$ & $.0025\pm.0005$ & $.0050\pm.0010$ & $.4636\pm.0222$ & $.8707\pm.0172$\\
\bottomrule
\end{tabularx}
\end{table}

The large gap between PR-AUC and recall for E1.2, E1.3, and E1.5 is consistent
with an operating-point transfer problem.  E1.4 raises recall but with very high
variance.  These numbers should not be compared to E1 as if the data
distribution and prevalence were unchanged.

\section{Expanded E5 Seed-Level Results}

\begin{table}[h]
\centering
\caption{E5 chronological fixed-threshold results by seed.}
\label{tab:e5seeds}
\small
\begin{tabular}{@{}rrrrrrr@{}}
\toprule
\textbf{Seed} & \textbf{Accuracy} & \textbf{MCC} & \textbf{Recall} &
\textbf{FPR} & \textbf{PR-AUC} & \textbf{ROC-AUC}\\
\midrule
13 & .965568 & -.002400 & 0 & .000168 & .090863 & .767225\\
21 & .965730 & 0 & 0 & 0 & .131614 & .795814\\
42 & .965016 & -.005035 & 0 & .000739 & .106793 & .784720\\
87 & .965730 & 0 & 0 & 0 & .112494 & .783162\\
101 & .965665 & -.001518 & 0 & .000067 & .106502 & .779585\\
\bottomrule
\end{tabular}
\end{table}

\section{E4 Reproduction Recipe}

From the root of the supplementary package, run:

\begin{lstlisting}[basicstyle=\ttfamily\small,backgroundcolor=\color{lightgray},frame=single]
python scripts/regenerate_e4.py \
  --rater-a analysis_inputs/E4_RaterA_scores.csv \
  --rater-b analysis_inputs/E4_RaterB_scores.csv \
  --output derived
\end{lstlisting}

The script asserts 200 rows per rater, 50 unique base cases, four conditions,
unique case--condition keys, and identical keys between raters.  It then writes
the merged rater-averaged data, condition means, all 15 Wilcoxon/Holm results,
bootstrap intervals, agreement statistics, and analysis-input SHA-256 hashes.

\FloatBarrier


\begin{thebibliography}{29}

\bibitem{ref1}
Kalouptsoglou, I., et al.  LocVul: Line-level vulnerability localization based
on a Sequence-to-Sequence approach.  \emph{Information and Software Technology}
189, 107940 (2026). \url{https://doi.org/10.1016/j.infsof.2025.107940}

\bibitem{ref2}
Ding, Y., Fu, Y., Ibrahim, O., Sitawarin, C., Chen, X., Alomair, B., Wagner,
D., Ray, B., Chen, Y.  Vulnerability Detection with Code Language Models: How
Far Are We?  In: \emph{ICSE}, 1729--1741 (2025).
\url{https://doi.org/10.1109/ICSE55347.2025.00038}

\bibitem{ref3}
Guo, D., Lu, S., Duan, N., Wang, Y., Zhou, M., Yin, J.  UniXCoder: Unified
Cross-Modal Pre-training for Code Representation.  In: \emph{ACL}, 7212--7225
(2022). \url{https://doi.org/10.18653/v1/2022.acl-long.499}

\bibitem{ref4}
Wang, Y., Wang, W., Joty, S., Hoi, S.C.H.  CodeT5: Identifier-aware Unified
Pre-trained Encoder-Decoder Models for Code Understanding and Generation.  In:
\emph{EMNLP}, 8696--8708 (2021).
\url{https://doi.org/10.18653/v1/2021.emnlp-main.685}

\bibitem{ref5}
Fu, M., Tantithamthavorn, C.  LineVul: A Transformer-based Line-Level
Vulnerability Prediction.  In: \emph{MSR}, 608--620 (2022).
\url{https://doi.org/10.1145/3524842.3528452}

\bibitem{ref6}
Fan, J., Li, Y., Wang, S., Nguyen, T.N.  A C/C++ Code Vulnerability Dataset with
Code Changes and CVE Summaries.  In: \emph{MSR}, 508--512 (2020).
\url{https://doi.org/10.1145/3379597.3387501}

\bibitem{ref7}
Hui, B., et al.  Qwen2.5-Coder Technical Report. arXiv:2409.12186 (2024).
\url{https://arxiv.org/abs/2409.12186}

\bibitem{ref8}
Mao, Q., Li, Z., Hu, X., Liu, K., Xia, X., Sun, J.  Towards Explainable
Vulnerability Detection with Large Language Models.  \emph{IEEE Transactions on
Software Engineering} 51(10), 2957--2971 (2025).
\url{https://doi.org/10.1109/TSE.2025.3605442}

\bibitem{ref9}
Du, X., Zheng, G., Wang, K., et al.  Vul-RAG: Enhancing LLM-based Vulnerability
Detection via Knowledge-level RAG.  \emph{ACM Transactions on Software
Engineering and Methodology} (2026).
\url{https://doi.org/10.1145/3797277}

\bibitem{ref10}
Weyssow, M., et al.  R2Vul: Learning to Reason about Software Vulnerabilities
with Reinforcement Learning and Structured Reasoning Distillation.
arXiv:2504.04699 (2025). \url{https://arxiv.org/abs/2504.04699}

\bibitem{ref11}
Zhu, H., Li, J., Gao, C., Qian, J., Dong, Y., Liu, H., Wang, L., Wang, Z., Hu,
X., Li, G.  VulInstruct: Teaching LLMs Root-Cause Reasoning for Vulnerability
Detection via Security Specifications.  \emph{Proceedings of the ACM on
Software Engineering} 3(FSE), 2813--2834 (2026).
\url{https://doi.org/10.1145/3808132}

\bibitem{ref12}
Lekssays, A., Mouhcine, H., Tran, K., Yu, T., Khalil, I.  LLMxCPG:
Context-Aware Vulnerability Detection Through Code Property Graph-Guided Large
Language Models.  In: \emph{USENIX Security}, 489--507 (2025).
\url{https://www.usenix.org/conference/usenixsecurity25/presentation/lekssays}

\bibitem{ref13}
Feng, Z., et al.  CodeBERT: A Pre-Trained Model for Programming and Natural
Languages.  In: \emph{Findings of EMNLP}, 1536--1547 (2020).
\url{https://doi.org/10.18653/v1/2020.findings-emnlp.139}

\bibitem{ref14}
Zhou, Y., Liu, S., Siow, J., Du, X., Liu, Y.  Devign: Effective Vulnerability
Identification by Learning Comprehensive Program Semantics via Graph Neural
Networks.  In: \emph{NeurIPS} 32 (2019).
\url{https://proceedings.neurips.cc/paper/2019/hash/49265d2447bc3bbfe9e76306ce40a31f-Abstract.html}

\bibitem{ref15}
Chen, Y., Ding, Z., Alowain, L., Chen, X., Wagner, D.  DiverseVul: A New
Vulnerable Source Code Dataset for Deep Learning Based Vulnerability Detection.
In: \emph{RAID} (2023). \url{https://doi.org/10.1145/3607199.3607242}

\bibitem{ref16}
Pintore, M., Piras, G., Sotgiu, A., et al.  Evaluating Line-level Localization
Ability of Learning-based Code Vulnerability Detection Models.  \emph{Machine
Learning} 115, 94 (2026). \url{https://doi.org/10.1007/s10994-025-06902-1}

\bibitem{ref17}
Guo, C., Pleiss, G., Sun, Y., Weinberger, K.Q.  On Calibration of Modern Neural
Networks.  In: \emph{ICML}, PMLR 70, 1321--1330 (2017).
\url{https://proceedings.mlr.press/v70/guo17a.html}

\bibitem{ref18}
Caruana, R.  Multitask Learning.  \emph{Machine Learning} 28, 41--75 (1997).
\url{https://doi.org/10.1023/A:1007379606734}

\bibitem{ref19}
Saito, T., Rehmsmeier, M.  The Precision-Recall Plot Is More Informative than
the ROC Plot When Evaluating Binary Classifiers on Imbalanced Datasets.
\emph{PLOS ONE} 10(3), e0118432 (2015).
\url{https://doi.org/10.1371/journal.pone.0118432}

\bibitem{ref20}
Cohen, J.  A Coefficient of Agreement for Nominal Scales.  \emph{Educational
and Psychological Measurement} 20(1), 37--46 (1960).
\url{https://doi.org/10.1177/001316446002000104}

\bibitem{ref21}
Kottner, J., et al.  Guidelines for Reporting Reliability and Agreement Studies
(GRRAS) Were Proposed.  \emph{Journal of Clinical Epidemiology} 64(1), 96--106
(2011). \url{https://doi.org/10.1016/j.jclinepi.2010.03.002}

\bibitem{ref22}
Nie, Y., Li, H., Guo, C., Jiang, R., Wang, Z., Li, B., Song, D., Guo, W.
VulnLLM-R: Specialized Reasoning LLM with Agent Scaffold for Vulnerability
Detection. arXiv:2512.07533 (2025).
\url{https://arxiv.org/abs/2512.07533}

\bibitem{ref23}
Hin, D., Kan, A., Chen, H., Babar, M.A.  LineVD: Statement-level Vulnerability
Detection Using Graph Neural Networks.  In: \emph{MSR}, 596--607 (2022).
\url{https://doi.org/10.1145/3524842.3527949}

\bibitem{ref24}
Yang, A.Z.H., Le Goues, C., Martins, R., Hellendoorn, V.J.  Large Language
Models for Test-Free Fault Localization.  In: \emph{ICSE}, 1--12 (2024).
\url{https://doi.org/10.1145/3597503.3623342}

\bibitem{ref25}
Pornprasit, C., Tantithamthavorn, C.  DeepLineDP: Towards a Deep Learning
Approach for Line-Level Defect Prediction.  \emph{IEEE Transactions on Software
Engineering} 49(1), 84--98 (2023).
\url{https://doi.org/10.1109/TSE.2022.3144348}

\bibitem{ref26}
Liu, E.T.S., Wang, A., Mateega, S., Georgescu, C., Tang, D.  VADER: A
Human-Evaluated Benchmark for Vulnerability Assessment, Detection, Explanation,
and Remediation. arXiv:2505.19395 (2025).
\url{https://arxiv.org/abs/2505.19395}

\bibitem{ref27}
He, J., Shi, J., Zhuo, T.Y., Treude, C., Sun, J., Xing, Z., Du, X., Lo, D.
LLM-as-a-Judge for Software Engineering: Literature Review, Vision, and the Road
Ahead. arXiv:2510.24367 (2025).
\url{https://arxiv.org/abs/2510.24367}

\bibitem{ref28}
Baltes, S., Angermeir, F., Arora, C., Mu\~noz Bar\'on, M., Chen, C., et al.
Guidelines for Empirical Studies in Software Engineering Involving Large
Language Models. arXiv:2508.15503 (2025).
\url{https://arxiv.org/abs/2508.15503}

\bibitem{ref29}
Le, H.-P.-L., Huynh, Q.-T.  X-LocVul: Reproducibility Artifact and Data for
Vulnerability Detection, Localization, and Explanation.  Version 1.0.0,
Zenodo (2026). \url{https://doi.org/10.5281/zenodo.22723493}

\end{thebibliography}
\end{document}